# Realizing Thermoelectric and Thermistor Bi-functionalities via Triggering Electron Correlations with Lattice-dipole

*Jikun Chen[1-3]†\*, Qihao Zhang[2]†, Jiaou Wang[4], Jie Xiao[2], Xinyou Ke[5], Takeaki Yajima[3], Feng Hao[2], Hongyi Chen[2], Yong Jiang[1]\*, Nuofu Chen[6] and Lidong Chen[2]\**

[1]Beijing Advanced Innovation Center for Materials Genome Engineering, School of Materials Science and Engineering, University of Science and Technology Beijing, Beijing 100083, China
[2]State Key Laboratory of High Performance Ceramics and Superfine Microstructure, Shanghai Institute of Ceramics, Chinese Academy of Sciences, Shanghai 200050, China
[3]School of Engineering, The University of Tokyo, Tokyo 1138656, Japan
[4]Beijing Synchrotron Radiation Facility, Institute of High Energy Physics, Chinese Academy of Sciences, Beijing 100049, China
[5]John A. Paulson School of Engineering and Applied Sciences, Harvard University, Cambridge, Massachusetts 02138, USA
[6]School of Renewable Energy, North China Electric Power University, Beijing 102206, China

Correspondences: Prof. Lidong Chen (cld@mail.sic.ac.cn), Prof. Yong Jiang (yjiang@ustb.edu.cn) and Prof. Jikun Chen (jikunchen@ustb.edu.cn). Request for materials: Prof. Jikun Chen (jikunchen@ustb.edu.cn).
†JC and QZ contribute equally to the present work.

**Abstract**

Establishing strong electron-correlations not only shed lights on overcoming the trade-off limitations for optimizing thermoelectric materials, but can also introduce new functionalities that extend the vision of conventional thermoelectric applications. Here, we demonstrate that the high thermoelectric and thermistor functionalities coexist in lattice distorted $SrNb_xTi_{1-x}O_3$ films with electron correlations between carriers and ordering aligned lattice dipoles. As-grown $SrNb_xTi_{1-x}O_3/SrTiO_3$ with effectively preserved interfacial strains exhibits cross-plane charge ordering and orbital anisotropy, as indicated by the polarization dependent near edge X-ray absorption fine structures. The resultant coulomb-correlations regulate the carrier transport and enhance the Seebeck coefficient more independently via enlarging the system's vibration entropy. As-achieved maximum thermoelectric power factor exceeds 100 $\mu Wcm^{-1}K^{-2}$ measured in the bulk performance of $SrNb_{0.2}Ti_{0.8}O_3$ (2.2 μm)/$SrTiO_3$ (100 μm), which is comparable to the best thermoelectric materials for low temperature applications. In addition, the strong temperature dependence of the carrier scattering aroused by the lattice dipoles introduces a positive temperature-dependent thermistor transportation behavior with large temperature coefficient of resistance ranging from 30 to 300 K, which is rarely seen in conventional thermistors. Combining both functionalities largely extend the horizon in exploring new Joule sensors for detection of temperature and thermal perturbations across a broad temperature range.

**Introduction**

Thermoelectric technique can achieve conversions between thermal and electrical energies based on miniaturized and compactly assembled devices, without involving conventional compressors and any mechanical movement.[1-5] It plays an irreplaceable role in two major aspects: localized refrigeration for microprocessors, and micro-scale power generations from environmental heat.[1,2] The thermoelectric energy conversion efficiency is determined by a figure of merit, $zT=S^2\sigma T\kappa^{-1}$, of thermoelectric materials ($S$: Seebeck coefficient, $\sigma$: electrical conductivity, $T$: absolute temperature, and $\kappa$: thermal conductivity).[3,4] Improving $zT$ should follow two directions: reducing lattice contribution in $\kappa$, or enhancing thermoelectric power factor ($PF=S^2\sigma$). For Future investigations, it is more urgent and required to achieve further enhancement in $PF$, since the lattice contribution to $\kappa$ for many thermoelectric material systems has been reduced towards the amorphous limit within the past two decades [3-5]. Although $PF$ fundamentally drives the thermoelectric conversion, it is bottlenecked by the reversely varied $\sigma$ and $S$ with the carrier concentration ($n$).[3] How to unlock this conventional $S$-$\sigma$ trade-off is one of the most vital issues to be addressed in the field of thermoelectric. Apart from improving $zT$ form the materials aspect, seeking for new applications of the thermoelectric technology is of equal importance. The horizon of thermoelectric will be extended beyond energy conversions, only if additional functionalities can be introduced into the thermoelectric materials to combine with the Seebeck or Peltier effects for achieving new device applications.

Triggering the electron correlations between carriers and ordering aligned array of lattice-dipoles sheds a light on realizing these concepts by breaking through the bottleneck in $PF$ [6-18] and meanwhile enriching thermoelectric materials with more functionalities. The past century witnesses the distinct characters discovered in the solid-state matters, such as metal-insulator transitions[9-13], superconductivities [14] and bad-metal conductors [15], via introducing un-negligible coulomb interactions among carriers-carriers and/or carriers-phonons. This was also the case for thermoelectric materials, as we noticed that the electron-phonon coupling (EPC) can cause additional changes in the system entropy associated to the thermal diffusion of carrier, and thereby contribute to an additional Seebeck coefficient ($S_{EPC}$) that is less dependent on $n$ or $\sigma$ [6-8, 16,17]. One typical example was indicated by an extraordinarily large and $n$-irrelevant $S$ (~300 μVK$^{-1}$) observed in the heavily doped boron carbides ($B_{12+x}C_{3-x}$).[6-8] The large magnitude of $S$ was mainly contributed by the extra change of entropy when localized carriers coupled with the singlet lattice-dipoles and softened the lattice vibration modes during carrier transport.[6] Analogous effects were also observed for the two-dimensional electron gas (2DEG) within a LaAlO$_3$/SrTiO$_3$ interfaces [16,17] or lattice distorted SrNb$_{0.2}$Ti$_{0.8}$O$_3$/SrTiO$_3$ [18], in which cases a large $S$ of ~10$^2$ μVK$^{-1}$ and a high $\sigma$ of ~10$^5$ Sm$^{-1}$ coexist as previously attributed to the electron phonon coupling (EPC). These demonstrations indicate an alternative direction to break the $S$-$\sigma$ trade-off via actively establishing the EPC, but this has not yet been fully explored compared with the other fields in solid-state matters. Apart from varying the system entropy, the presence of EPC is also expected to strengthen the temperature dependence of the electrical transportations and enhance the temperature coefficient of resistance (TCR). As a result, an additional functionality of the thermistor transport will be introduced into the thermoelectric material, creating a combined functionality of thermoelectric and thermistor.

In this work, we demonstrate the co-existence of high thermoelectric and thermistor transport performances achieved in lattice distorted $SrNb_xTi_{1-x}O_3$, by triggering the electron correlations between carriers and the ordering aligned lattice dipoles. The distinct electronic structures of cross-plane ordering and orbital anisotropy were indicated by the polarization dependant near edge X-ray absorption fine structure (NEXAFS). By controlling the lattice mismatch and deposition thickness, we regulated the magnitude of the preserved lattice distortion that is imparted upon $SrNb_xTi_{1-x}O_3$, and further investigated their relationships with both the thermoelectric and thermistor performances. Potential applications as a Joule sensor for detections of temperature and thermal perturbation are further explored based on the combined functionality of the thermoelectric and thermistor.

**Concept**

Figure 1a illustrates the general concept to establish an electron - lattice dipole correlated system within niobium doped strontium titanate when binding the distortions induced lattice polarization and interfacial polarization. The strontium titanate is known to exhibit a ferroelectric nature at room temperature when distorted by interfacial strain.[18-22] Accordingly, a coulomb correlation between carrier and lattice-diploes will be established by coherently depositing $SrNb_xTi_{1-x}O_3$ on single-crystal $SrTiO_3$ substrates under appropriate deposition kinetics. With a larger intrinsic lattice constant than the $SrTiO_3$ substrate, the co-lattice grown $SrNb_xTi_{1-x}O_3$ film is under biaxial in-plane compressive distortion, and this is known to separate the charge center of the $TiO_6$ octahedra and form cross-plane $Ti^+ \rightarrow O^-$ lattice-dipoles.[22] Thermodynamically, as formed lattice-dipoles favored to bind with the interfacial polarization to reach an ordering alignment ($\vec{O}$) in cross-plane.[18,27,28] The resultant periodical coulomb potential wells is expected to regulate the carrier distribution and transportations and result in the following two effects.

**1) Overcoming $S$-$\sigma$ trade-off and improving $PF$:** The $S$ mainly originates form the derivative in the system entropy versus the number of carriers. For conventional thermoelectric materials, $S$ is contributed by the variations in entropy-of-mixing by changing the carrier occupation rate in electronic states ($c$).[8] The resultant Seebeck coefficient resulting for entropy-of-mixing was determined by $S_{mix} = (\frac{k_B}{e}) \ln \frac{1-c}{c}$, where $k_B$ is the Boltzmann constant and $c$ represents the occupation rate of the energy state near the Fermi energy.[6] Apparently, $S_{mix}$ will increase by reducing carrier concentration ($n$) and an enhancing density of states (DOS) near the Fermi level[3], since both ways reduces $c$ and enlarge the entropy change associated to carrier transport driven by temperature gradient. This is also the origin of the conventional $S$-$\sigma$ trade-off, since elevating the doping level increases $n$ and $\sigma$, but meanwhile reduces $S_{mix}$.

By establishing correlations with the lattice-dipoles, per carrier transport induced variation in the system's entropy can be further enlarged via altering the lattice vibration [6,7], contributing an extra $S_{EPC}$ to the Seebeck coefficient ($S = S_{EPC} + S_{mix}$). Treating the vibrations of lattice dipoles are treated as harmonic vibrations at small vibration amplitudes, the $S_{EPC}$ is described as [6,7]:

$$S_{EPC} = S_{Vabriation} + S_{Transport} = (\frac{k_B}{e})\sum_i \{\frac{-\Delta\omega_i}{\omega_i}[\frac{\hbar\omega_i/2k_BT}{\sinh(\hbar\omega_i/2k_BT)}]^2 + \frac{E_{A,i}\hbar\omega_i/2k_BT}{\sinh(\hbar\omega_i/2k_BT)}\} \quad (1)$$

Where, the $\Delta\omega_i$ represents the changes in vibration frequency associated to coupling with carriers at each vibration frequency of $\omega_i$, $E_{A,i}$ is the activation energy for carrier transportation at $\omega_i$, $k_B$ is the Boltzmann constant and $\hbar$ is the reduced Planck constant. As indicated by Eq. (1), $S_{EPC}$ is mainly determined by the number of coupled frequency between electron and phonons, and the respective magnitude of $\Delta\omega_i$ and $E_{A,i}$ at each frequency. This indicates the two keys to achieve large $S_{EPC}$: strengthening the strain-induced lattice polarization and enhancing their couplings.

**2) Introducing thermistor transportations with large TCR:** The vibrations of lattice dipoles are usually activated by phonons, and their further coupling with the carrier transport is expected to strengthen the temperature dependence of the resistivity. For example, elevating the temperature will promote the vibrations of the lattice dipoles that result in more significantly carrier scatterings, as compared to conventional semiconductors. Therefore, a more dramatic enhancement in the resistivity of the material is expected, achieving a thermistor transportation behavior with a relatively large positive TCR. This concept was confirmed by the observed large positive temperature dependence in the 2DEGs within $LaAlO_3/SrTiO_3$ interfaces [16,17], in which case the electron carriers underneath the interface correlated with the interfacial polarizations that are perpendicular to the carrier transport. Compared with the $LaAlO_3/SrTiO_3$, as-proposed lattice distorted $SrNb_xTi_{1-x}O_3/SrTiO_3$ exhibits a similar electronic structure and the way in carrier transportation. For example, interfacial polarization at $SrNb_xTi_{1-x}O_3/SrTiO_3$ can be considered to be effectively extended within the $SrNb_xTi_{1-x}O_3$ film materials, resulting in a thermistor transportation behavior.

The thermistor transport functionality was applied previously in aspects such as a Joule sensor for detections of temperature and/or thermal perturbation, circuit protection and current limiting, and temperature compensated synthesizer voltage controlled oscillators [39-41]. Nevertheless, one challenge in Joule sensing is the relative small magnitude of TCR (i.e. ∼1-2 % K$^{-1}$) of conventional thermistors, resulting in relatively small magnitude of the reading out voltage signal compared to the background signal introduced by the input electric current signal at a small temperature raise. Compared to Joule sensing via thermistor, the Seebeck effect in thermoelectric functionality provides another direction to sense the incident Joule heat via its resultant temperature difference. This process does utilize any input electronic signals that easily enhances the background signal, but requires a reaction time to establish a temperature gradient that further generates the Seebeck voltage to be read out. Therefore, a larger tolerance in compromising between high detection resolution and fast reaction speed is expected in a new Joule sensor combining both the thermoelectric and thermistor functionalities.

**Results and discussions**

To achieve as-proposed concept and further regulate both thermoelectric and thermistor transport functionalities, we adjust the lattice constant ($a_0$) and deposition thickness ($t_{Film}$) of the films growing on SrTiO$_3$ (001) substrate to control the preservation of interfacial strain and its induced lattice polarization. For example, single-layer SrNb$_{0.2}$Ti$_{0.8}$O$_3$ or SrNb$_{0.4}$Ti$_{0.6}$O$_3$ were deposited at various $t_{Film}$ on SrTiO$_3$ (001). In addition, laminated layers, such as SrNb$_{0.4}$Ti$_{0.6}$O$_3$ /SrNb$_{0.2}$Ti$_{0.8}$O$_3$ and SrNb$_{0.4}$Ti$_{0.6}$O$_3$ /SrNb$_{0.2}$Ti$_{0.8}$O$_3$, at a similar $t_{Film}$ were also deposited on SrTiO$_3$ (001). The SrNb$_{0.2}$Ti$_{0.8}$O$_3$ ($a_0$= 3.96 Å) exhibits a smaller lattice mismatch ($\varepsilon$=1.42%) with the SrTiO$_3$ ($a_0$= 3.905 Å) substrate, compared to the ones for SrNb$_{0.4}$Ti$_{0.6}$O$_3$ ($a_0$= 4.00 Å) with SrTiO$_3$ ($\varepsilon$=2.55 %). As a result, a better coherency in the epitaxy was observed when depositing SrNb$_{0.2}$Ti$_{0.8}$O$_3$ compared with SrNb$_{0.4}$Ti$_{0.6}$O$_3$. This is further indicated by the reflection high-energy electron diffraction results (see Figure S1 with more discussions in SI: section C).

The lattice distortions of the film material are quantified from reciprocal space mappings (RSM), as shown in Figure 1b for four representative films at similar deposition thicknesses. Their X-ray diffraction (XRD) patterns are demonstrated in Figure S2. In the RSM spectrums, $I_{Film(Q_{//},Q_{\perp})}$ and $I_{Substr.(Q_{//},Q_{\perp})}$, represent the material diffractions distributed at the in-plane ($Q_{//}$) and cross-plane ($Q_{\perp}$) reciprocal coordinate, respectively. Accordingly, the averaged biaxial lattice distortion and the percentage of materials distributed in each magnitude of lattice distortion were calculated[18]. For single layered films at a similar deposition thickness, the lattice compression is more peserved for the SrNb$_{0.2}$Ti$_{0.8}$O$_3$ /SrTiO$_3$, compared to SrNb$_{0.4}$Ti$_{0.6}$O$_3$ /SrTiO$_3$ (see the strain distributions in Figure 1b). For laminated layered films, the sequence of layer lamination can also influence the preservation of lattice distortions, since the strain relaxes differently at various $\varepsilon$. This is seen from the better preserved lattice distortion for SrNb$_{0.4}$Ti$_{0.6}$O$_3$ /SrNb$_{0.2}$Ti$_{0.8}$O$_3$ /SrTiO$_3$, compared to SrNb$_{0.4}$Ti$_{0.6}$O$_3$ /SrNb$_{0.2}$Ti$_{0.8}$O$_3$ /SrTiO$_3$.

The interfacial strain induce compressive lattice distortion is expected to further influence the electronic structures and charge ordering of as-grown film material. Therefore, the polarization dependent near edge X-ray absorption fine structures (NEXAFS)[29-31] were performed for SrNb$_{0.2}$Ti$_{0.8}$O$_3$ (400 nm)/SrTiO$_3$, SrNb$_{0.2}$Ti$_{0.8}$O$_3$ (400 nm)/LaAlO$_3$, SrNb$_{0.4}$Ti$_{0.6}$O$_3$ (400 nm)/SrTiO$_3$ and SrTiO$_3$ single crystal, as the results shown in Figure 1c. Since the depth in response of the X-ray absorption was around 100 nm, as-obtained signal is from the film material.[29] The conduction band splits into four energy levels, among which the two lower energy orbits of $t_{2g}$ and $e_g$ originate from O-2$p$ hybridized with Ti-3$d$ orbits.[32-34] The undistorted or strain relaxed SrTiO$_3$ (or SrNb$_x$Ti$_{1-x}$O$_3$) exhibits an octahedral symmetry (TiO$_6$), in which case the $e_g$ orbits from hybridization between O-2$p$ and Ti-3$d$ are expected to be isotropy for the in-plane and cross-plane directions. This is in agreement with the non-angular dependent NEXAFS spectrums observed for the undistorted $SrTiO_3$ single crystals (Figure S3), the strain-relaxed $SrNb_{0.2}Ti_{0.8}O_3/LaAlO_3$ (Figure 1c, middle), and partially strain-relaxed and $SrNb_{0.4}Ti_{0.6}O_3/SrTiO_3$ (Figure 1c, right).

In contrast, a polarization dependent $e_g$ orbit is observed for $SrNb_{0.2}Ti_{0.8}O_3$ /$SrTiO_3$ (Figure 1c, left), indicating a strong orbital anisotropy in the $\sigma$-bond hybridized by O-$2p$ ($\sigma$) hybridized with Ti-$3d$. An enhanced resonation is observed when the electric field from the polarized X-ray exhibits smaller angle with respective to the cross-plane direction. This is in consistency with the resonance of electrons in the dipole approximation with the linearly polarized X-ray from synchrotron radiations, as described previously to be [29]: $I \propto |\langle \psi_f | e \bullet p | \psi_i \rangle|^2 \propto |\vec{E} \bullet \vec{O}|^2$. The term $|\langle \psi_f | e \bullet p | \psi_i \rangle|^2$ determines the magnitude of resonance cross-section and is proportional to the absorption intensity ($I$), where $\Psi_f$ and $\Psi_f$ represent the final and initial state, $e$ is the unit electric field vector, and $p$ is the dipole transition operator.

Similar phenomenon was previously observed in ferroelectric or magnetic domains of materials, such as $BiFeO_3$[30] and $FeTiO_3$[31]. For the present $SrNb_{0.2}Ti_{0.8}O_3$ /$SrTiO_3$, the aligned $Ti^+ \rightarrow O^-$ lattice-dipoles and their influence in the distribution of electron carriers along the cross-plane direction are expected to form an ordering and directional polarization state ($\vec{O}$). The resonance is maximized when $\vec{O}$ is along the electromagnetic filed of the polarized X-ray ($\vec{E}$) and minimized when $\vec{O}$ is perpendicular to $\vec{E}$.[29-32] It is also interesting to note that compared with $SrNb_{0.2}Ti_{0.8}O_3$ /$LaAlO_3$, the relative absorption intensity for the $e_g$ orbit is lower for $SrNb_{0.2}Ti_{0.8}O_3$ /$SrTiO_3$, in particular along the cross-plane direction. This indicates a reduced proportion of DOS hybridized from O-$2p$ in the conduction band, and should be associated to the enhanced coulomb repulsion between the lattice-dipoles regulated in-plane electron carriers with their adjacent oxygen sites.

The orderly alignment in the charge anisotropic lattice-dipoles in $SrNb_{0.2}Ti_{0.8}O_3/SrTiO_3$ is expected to zigzag the previous conduction band owning to their respective impact on the carrier distributions and transportations [15,16,18,24]. In Figure 2a, the $S$ and deposition induced sheet conductance ($\sigma_{xx}/t_{Film}$) measured at room temperature are compared for as-grown $SrNb_xTi_{1-x}O_3$ at various magnitudes of lattice distortions. In general, both the $S$ and $\sigma_{xx}/t_{Film}$ for $SrNb_xTi_{1-x}O_3/SrTiO_3$ shows an increasing tendency with an enlarged lattice distortion. In contrast, the $SrNb_xTi_{1-x}O_3/LaAlO_3$ and $SrNb_xTi_{1-x}O_3/(LaAlO_3)_{0.3}(Sr_2AlTaO_6)_{0.7}$ with completely relaxed interfacial strains (see RSM spectrum in Figure S4) exhibits a similar $S$ and $\sigma_{xx}/t_{Film}$ to their respective bulk values at similar compositions. The deposition induced carrier density and carrier mobility were further shown in Figure S5 for several representative samples. When growing on $LaAlO_3$ substrate to relax the interfacial strains, a larger carrier density (Figure S5a) and smaller mobility (Figure S5b) is observed for $SrNb_{0.4}Ti_{0.6}O_3$ as compared to $SrNb_{0.2}Ti_{0.8}O_3$. This tendency is in agreement to the conventional thermoelectric bulk materials, when increasing the doping concentration. When growing on the $SrTiO_3$ substrate, both the carrier density (Figure 5c) and mobility (Figure 5d) enhances with the magnitude of the preserved lattice distortions. This observation reveals that the carrier formation and transportation for the presently grown $SrNb_xTi_{1-x}O_3/SrTiO_3$ is more dominated by the lattice distortion or polarization rather than element doping, and this differs from conventional thermoelectric materials.

The above understanding is further supported by comparing the temperature dependence in the electrical transportation properties. As shown Figure 2b, the sheet resistance ($R_{xx}$) for the strain relaxed $SrNb_xTi_{1-x}O_3/LaAlO_3$ remains nearly constant when varying the temperature ($T$), and this is similar to the tendency observed for heavily doped $SrNb_xTi_{1-x}O_3$ bulk materials. In contrast, the $SrNb_xTi_{1-x}O_3/SrTiO_3$, which exhibits a thermistor transportation behaviour with steeply increasing $R_{xx}$-$T$ tendencies across a broad range of temperature from 10 to 300 K. Elevating the temperature results in a reduction of their TCR reduces from a magnitudes of ~5 % $K^{-1}$ at 10 K to ~1 % $K^{-1}$ at 300 K. A further comparison indicates a larger TCR near room temperature is observed for more distorted samples, e.g. $SrNb_{0.2}Ti_{0.8}O_3/SrTiO_3$ compared to $SrNb_{0.4}Ti_{0.6}O_3/SrTiO_3$.

To better understand the $R_{xx}$-$T$ tendencies, Hall measurement was performed to investigate the temperature dependence of the sheet carrier density ($n_{xx}=R_{Hall}^{-1}$) and carrier mobility ($\mu_{Hall}$), as their results demonstrated in Figure 2c. A nearly constant $n_{xx}$ is observed for $SrNb_{0.2}Ti_{0.8}O_3/SrTiO_3$, $SrNb_{0.4}Ti_{0.6}O_3/SrNb_{0.2}Ti_{0.8}O_3/SrTiO_3$, and $SrNb_{0.2}Ti_{0.8}O_3/LaAlO_3$, in which situations the interfacial strain either preserved or completely relaxed. For strain partially relaxed samples, such as $SrNb_{0.4}Ti_{0.6}O_3/SrTiO_3$ and $SrNb_{0.2}Ti_{0.8}O_3/SrNb_{0.4}Ti_{0.6}O_3/SrTiO_3$, elevating $T$ results in a more significant enhancement in their $n_{xx}$. This may associate to the thermal activation of carriers that were previously trapped by the lattice defects within the strain gradual relaxation region. The $\mu_{Hall}$-$T$ observed for $SrNb_xTi_{1-x}O_3/SrTiO_3$ exhibits a significantly reducing tendencies similar to the previously reported ones for the 2DEGs within the $LaAlO_3/SrTiO_3$ interfaces [16,17], and in contrast to the constant $\mu_{Hall}$–$T$ tendency observed for $SrNb_{0.2}Ti_{0.8}O_3/LaAlO_3$.

To further cater to practical applications in thermoelectric energy conversions, the deposition thickness ($t_{Film}$) was increased beyond a micrometer-scale, and meanwhile the thickness of the $SrTiO_3$ (001) substrates ($t_{Substrate}$) was reduced to 100 µm. Four representative $SrNb_xTi_{1-x}O_3$ (1.2-2.8 µm) /$SrTiO_3$ (100 µm) samples were prepared under the same deposition condition for the present investigation. Owing to a smaller lattice mismatch, as-deposited $SrNb_{0.2}Ti_{0.8}O_3$ /$SrTiO_3$ (sample A and B) and $SrNb_{0.1}Ti_{0.9}O_3$ /$SrTiO_3$ (sample D) are more distorted compared with the $SrNb_{0.4}Ti_{0.6}O_3$ /$SrTiO_3$ (sample C), as shown in Figure S6. Figure 3a shows the temperature dependent $S$ and $\sigma$ for four samples, measured as a bulk material by integrating the thickness of both films and substrates where $\sigma=\sigma_{xx} (t_{Film} + t_{Substrate})^{-1}$. At room temperature, sample C exhibits a similar Seebeck coefficient to the bulk value of $SrNb_{0.4}Ti_{0.6}O_3$, while its sheet conductance ($\sigma_{xx} = 0.47$ $\Omega^{-1}$ sq.) is similar to a 1.6 μm thick SrNb$_{0.4}$Ti$_{0.6}$O$_3$ bulk layer ($\sigma_{xx}$ = 0.41 Ω$^{-1}$ sq.). In contrast, the sheet resistance for sample A, B and D are around three times larger than the ones from their respective bulk layers at the same thicknesses. Compared with the strain-relaxed SrNb$_{0.4}$Ti$_{0.6}$O$_3$/SrTiO$_3$ (Sample C), the distorted SrNb$_{0.2}$Ti$_{0.8}$O$_3$/SrTiO$_3$ (Sample A and B) and SrNb$_{0.1}$Ti$_{0.9}$O$_3$/SrTiO$_3$ (Sample D) exhibit larger $S$ and $\sigma$ over all the measured temperature range of 10-300 K. These results are consistent with the ones observed in Figure 2a. Comparing the performance of Sample A and B with different $t_{Film}$, the sheet conductance shows a linear enhancement with the thin film deposition thickness as demonstrated in Figure S7. We can further estimate the additional introduced conductance and Seebeck coefficient by increasing the deposition thickness of 1.6 μm, via $\sigma_{1.6\mu m}=(\sigma_{xx,B}-\sigma_{xx,A})(t_B-t_A)^{-1}$ and $S_{1.6\mu m}=(S_B\sigma_{xx,B}-S_A\sigma_{xx,A})(\sigma_{xx,B}-\sigma_{xx,A})^{-1}$, respectively, as shown in Figure S8.

In Figure 3b, we further compare the thermoelectric power factor (*PF*) for the Sample A and B with the reported thermoelectric materials[35-38]. It can be seen that the presently achieved maximum *PF* when treating the SrNb$_{0.2}$Ti$_{0.8}$O$_3$ (2.8 μm)/SrTiO$_3$ (100 μm) sample as bulk material (by considering the sum of thicknesses for the film and substrate) is approaching to the one observed in Bi$_{1-x}$Sb$_x$ at low temperature range. The dash-line in Figure 3b shows the calculated magnitude for using $S_{1.6\mu m}^2\sigma_{1.6\mu m}$, indicating the introduced *PF* by depositing additional 1.6 μm film material. By either calculating the $\kappa_{Film}$ from as measured $\kappa_{Film\&Substrate}$ (Figure S9a) or performing similar estimation to ref [24], the obtained $\kappa_{Film}$ is in the magnitude of ~13-15 Wm$^{-1}$K$^{-1}$ (Figure S9b). More details of the two approaches for calculating the $\kappa_{Film}$ are demonstrated in the supporting information (Figure S9). The resultant *zT* values associated to the film materials at room temperature for sample A-C are estimated to be of ~0.8 to 1.0 (Figure S9c), a room temperature performance comparable to Bi$_2$Se$_{0.3}$Te$_{2.7}$[4]. By further reducing the temperature, the low temperature *zT* values achieved presently are expected to exceed the reported ones from the best low temperature thermoelectric materials such as Bi$_{1-x}$Sb$_x$[36] and CsBi$_4$Te$_6$[37] as demonstrated in Figure S9d.

The enhancement in thermoelectric performances is expected to be associated to the unlocked *S-σ* inter-relations when actively triggered carriers coulomb correlations with the orderly aligned cross-plane array of the lattice dipoles. In Figure 3c, the *S-σ* tendency at room temperature for samples in this work are compared with the reported ones for bulk or 2DEG-related SrTiO$_3$[24-26], as well as the typical thermoelectric materials used near room temperature or below.[35-37] The film materials of SrNb$_{0.2}$Ti$_{0.8}$O$_3$ and SrNb$_{0.1}$Ti$_{0.9}$O$_3$ in this work exhibit a large $S$ of ~ 320 μVK$^{-1}$ at a relatively high electrical conductivity of ~ 4.5 x 10$^5$ Sm$^{-1}$. Their respective *S-σ* relations at room temperature follow ones observed for the SrTiO$_3$-related 2DEGs.[24,25] In contrast, the *S-σ* relation of the strain-relaxed SrNb$_{0.4}$Ti$_{0.6}$O$_3$ film material follows the doped-SrTiO$_3$ as a bulk material. This is in agreement to the NEXAFS results demonstrated in Figure 1c, since an ordering aligned lattice polarization structure along the cross-plane direction is expected to form periodical well-like potentials. This results in more centralized spatial distribution of the carriers within the potential-well and forms a multi-channel quasi-2D transportations, similar to the ones observed for superlattice or interfacial 2DEGs.

Combining both functionalities of high thermoelectric and thermistor transportation can extend the vision in further exploring new applications of the thermoelectric materials apart from conventional thermoelectric energy conversions. One example demonstrated in Figure 4 is to achieve broad-range temperature sensing and imaging the localized thermal perturbation based on both functionalities of the presently grown $SrNb_xTi_{1-x}O_3/SrTiO_3$. Figure 4a demonstrates the flow chart of the detection strategy. The film material can be manufactured into the T-shape and aligned in an array, in which situation the thermistor and thermoelectric functionalities are used along the transverse and longitudinal ways, respectively. Under a given environment, the temperature can be detected via the $R_{xx}$-$T$ tendency, as one example given in Figure 4b and S10.

When knowing the temperature, the device can be further used for sensing or imaging the thermal perturbation via two modes: 1) the fast response mode based on the functionality of thermistor; 2) the high resolution based on the functionality of thermoelectric. We further performed the finite element simulation on the detection prototype to compare the two different modes in Joule detections via thermistor and thermoelectric functionalities, the results of which are detailed in Figure S10. As a thermal perturbation is applied onto the crossing point, the raised temperature ($\triangle T$) in localized region can be fast sensed via the enhancement in the transverse resistance ($R_T$). This can be detected by applying a transverse current $I_0$ and the change of transverse voltage ($V_R$) can be sensed as $V_R/V_{R0}= (1+\text{TCR}\triangle T)$. Meanwhile, the magnitude of $\triangle T$ can be better resolved via its resultant Seebeck voltage ($V_s=S\triangle T$) in longitudinal way via the thermoelectric functionality, utilizing the large magnitude of $S$ ($\sim 10^2$ μV K$^{-1}$). Figure 4c demonstrates the temperature dependence of the $TCR$ and $S$ for the present $SrNb_xTi_{1-x}O_3/SrTiO_3$ (100 μm) samples in comparison with the ones for BiSb, which is the most commonly used thermoelectric materials in the low temperature range. It can be seen that within a broad temperature range from 70 K to room temperature, both the TCR and $S$ of the $SrNb_xTi_{1-x}O_3/SrTiO_3$ far exceeds the ones observed for the conventional low temperature thermoelectric material of BiSb [36].

**Conclusion**

In summary, both functionalities of high thermoelectric and thermistor transportation were achieved in lattice distorted $SrNb_xTi_{1-x}O_3$ via triggering the coulomb correlations between the carrier and ordering aligned lattice dipoles. Binding the strain induced lattice polarization with the interfacial polarization results in ordering alignment of the cross-plane Ti$^+$→O$^-$ lattice-dipoles that couples with the carrier transport, as indicated by polarization dependent NEXAF analysis. As a result, the change of entropy from softening the phonon mode was enhanced that introduces an additional $S_{EPC}$ to break the conventional $S$-$\sigma$ trade-off and further improve the thermoelectric performances. Even considering the film and substrate as a bulk material, the maximum $PF$ of the $SrNb_{0.2}Ti_{0.8}O_3$ (2.8 μm)/SrTiO$_3$ (100 μm) sample reaches to ~ 103 μWcm$^{-1}$K$^{-2}$ at 45 K. This should be the highest $PF$ ever reported for thermoelectric oxides as bulk materials and approaches the ones for the best report from $Bi_{1-x}Sb_x$ at low temperatures. In addition to the thermoelectric functionality, a thermistor transportation behavior with large TCR across a broad temperature range was achieved in as-grown lattice distorted $SrNb_xTi_{1-x}O_3/SrTiO_3$, owning to the temperature dependence of the carrier scattering by the lattice dipoles. Combining both functionalities of thermoelectric and thermistor transportation will largely extend the vision in designing new electronic devices and exploring their new applications in aspects such as Joule sensor for detections of the temperature and thermal perturbations.

**Acknowledgments**
This work is supported by the National Natural Science Foundation of China (No. 51602022, No. 61674013 and No. 1157227), the key research program of Chinese Academy of Sciences (Grant No. KGZD-EW-T06) and the research grant from Shanghai government (No. 15JC1400301). JC also appreciate Japanese Society for the Promotion of Science (Fellowship ID: P15363). We appreciate helpful discussions and technical supports by Prof. Akira Toriumi from The University of Tokyo (Japan), Dr. Guolei Xiang from University of Cambridge (United Kingdom) and Prof. Renkui Zheng from Shanghai Institute of Ceramics, Chinese Academy of Sciences (China).
**Competing Interests:** The authors declare no competing financial interest.
**Additional Information:** Supplementary Information is available for this manuscript.
**Correspondences:** Correspondences should be addressed to: Prof. Lidong Chen (cld@mail.sic.ac.cn), Prof. Yong Jiang (yjiang@ustb.edu.cn) and Prof. Jikun Chen (jikunchen@ustb.edu.cn). Request for materials should contact Prof. Jikun Chen (jikunchen@ustb.edu.cn).
**Author Contributions:** JC conceived the idea, planed for the experiments, developed the film deposition strategy, performed film growth, and wrote the manuscript assisted by LC and QZ; QZ performed the simulation; JW and JC managed the synchrotron experiment and data analysis; JX, HC and FH characterized the transportation performance; JC and XK contributed to the RSM measurement and analysis; JC, YT, NC and YJ provided experimental supports and useful discussions.

**Figures and Captions:**

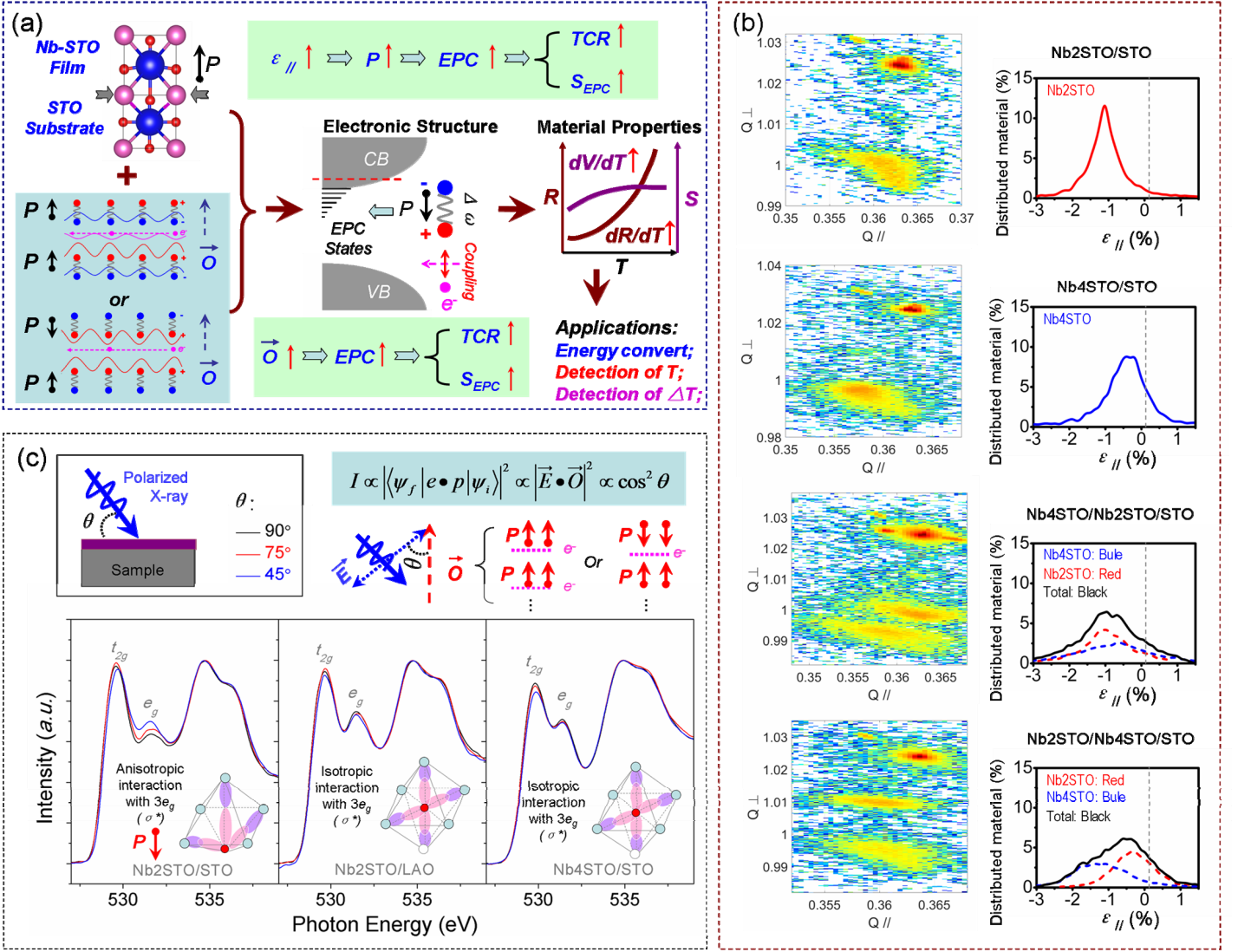

**Figure 1.** (a) Schematic illustration of a concept to enhance the Seebeck coefficient by coupling the cross-plane ordering aligned ferroelectric polarons and electron carriers, and meanwhile establishing the thermistor transportation functionality. (b) Reciprocal space mapping (RSM) and the percentage of material distributed as a function of the lattice distortions for: $SrNb_{0.2}Ti_{0.8}O_3$ ($t$= 440 nm) /$SrTiO_3$, $SrNb_{0.4}Ti_{0.6}O_3$ ($t$= 390 nm) /$SrTiO_3$, $SrNb_{0.4}Ti_{0.6}O_3$ ($t$= 200 nm) /$SrNb_{0.2}Ti_{0.8}O_3$ ($t$= 220 nm) /$SrTiO_3$, $SrNb_{0.2}Ti_{0.8}O_3$ ($t$= 220 nm) /$SrNb_{0.4}Ti_{0.6}O_3$ ($t$= 200 nm) /$SrTiO_3$ samples. (c) Polarization dependent near edge X-ray absorption fine structure (NEXAFS) analysis for $SrNb_{0.2}Ti_{0.8}O_3$ (400 nm)/$SrTiO_3$, $SrNb_{0.2}Ti_{0.8}O_3$ (400 nm)/$LaAlO_3$ and $SrNb_{0.4}Ti_{0.6}O_3$ (400 nm) /$SrTiO_3$.

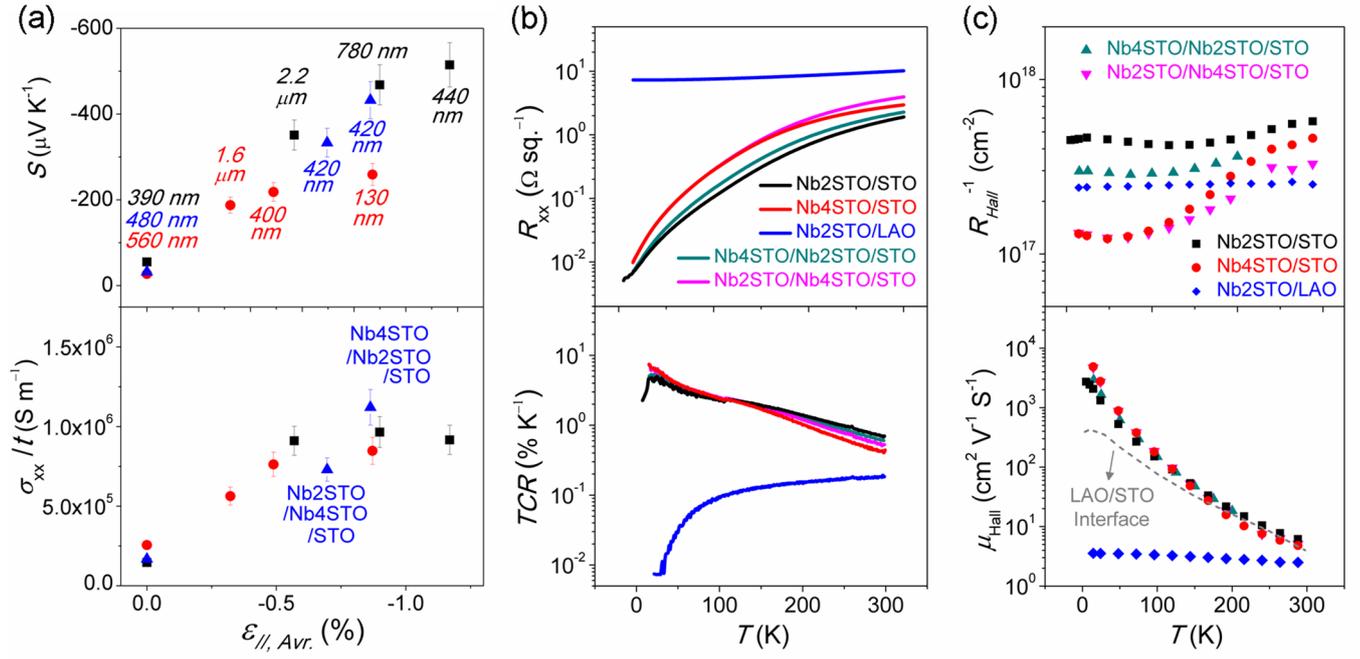

**Figure 2.** **(a)** Seebeck coefficient (*S*) and sheet conductance ($\sigma_{xx}$) divided by deposition thickness (*t*) for $SrNb_xTi_{1-x}O_3$ film samples at different lattice distortion ($\varepsilon_{//, Avr.}$). **(b)** The sheet resistance ($R_{xx}$) and temperature coefficient of resistance (*TCR*), **(c)** sheet carrier concentration ($R_{Hall}^{-1}$) and Hall mobility ($\mu_{Hall}$) for several representative $SrNb_xTi_{1-x}O_3$ film samples measured as a function of temperature from 5-300 K.

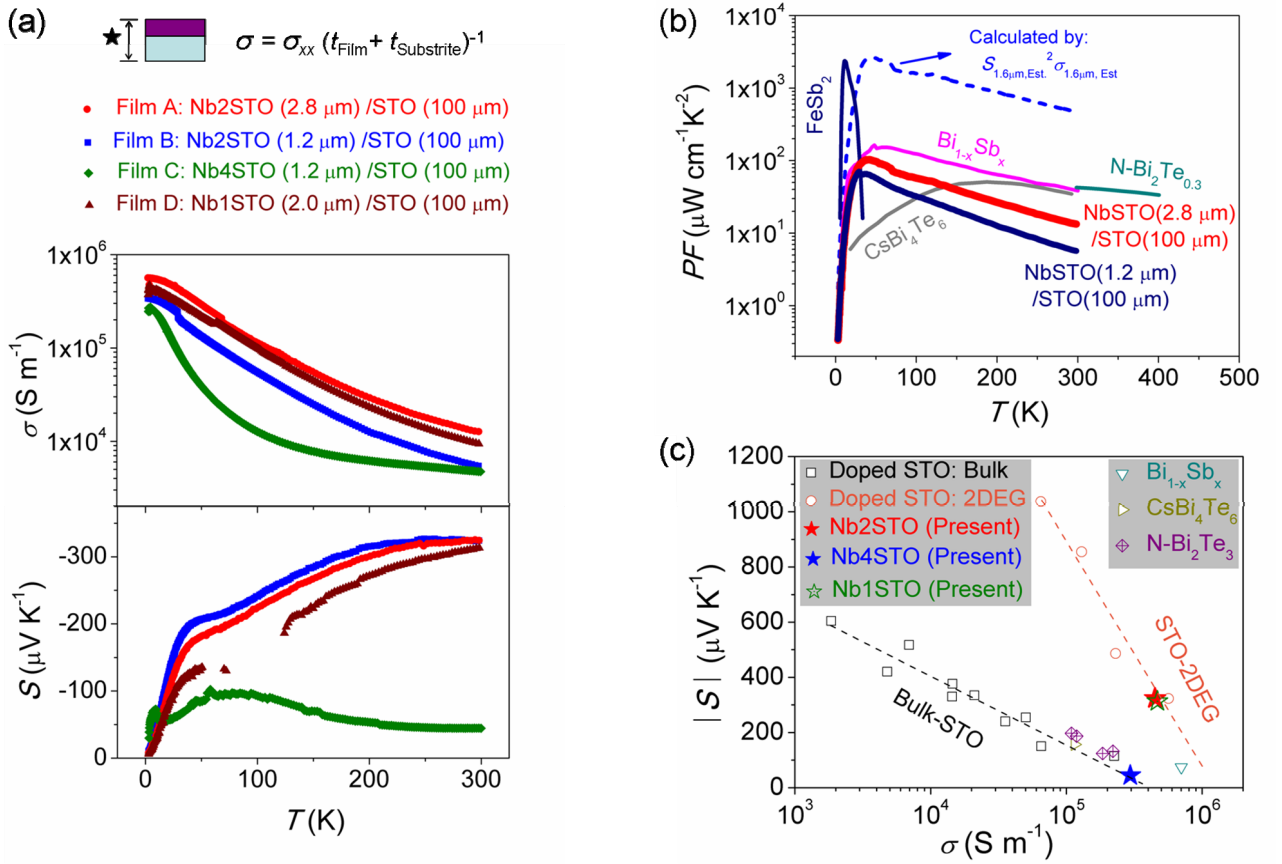

**Figure 3. (a)** Temperature dependence of electrical conductivity ($\sigma$) and Seebeck coefficient ($S$) for the following four samples: A. SrNb$_{0.2}$Ti$_{0.8}$O$_3$ ($t$= 1.2 μm) /SrTiO$_3$ ($t$= 100 μm), B. SrNb$_{0.2}$Ti$_{0.8}$O$_3$ ($t$= 2.8 μm) /SrTiO$_3$ ($t$= 100 μm), C. SrNb$_{0.4}$Ti$_{0.6}$O$_3$ ($t$= 1.6 μm) /SrTiO$_3$ ($t$= 100 μm), and D. SrNb$_{0.1}$Ti$_{0.9}$O$_3$ ($t$= 2 μm) /SrTiO$_3$ ($t$= 100 μm). As-shown $\sigma$ as is calculated by sheet conductivity divided by the sum-up thickness of the film and substrate (as a bulk material). **(b)** Thermoelectric power factor (*PF*) for sample A-D. Even taking the substrate into account, as achieved maximum *PF* for SrNb$_{0.2}$Ti$_{0.8}$O$_3$ (2.8 μm) /SrTiO$_3$ (100 μm) reaches to ~103 μWm$^{-1}$K$^{-2}$, which is the highest value among the oxide thermoelectric bulk material [35-37]. **(c)** *S*-$\sigma$ correlations for above samples compared with the ones reported for 2DEG and bulk thermoelectric materials [23-26, 35-37].

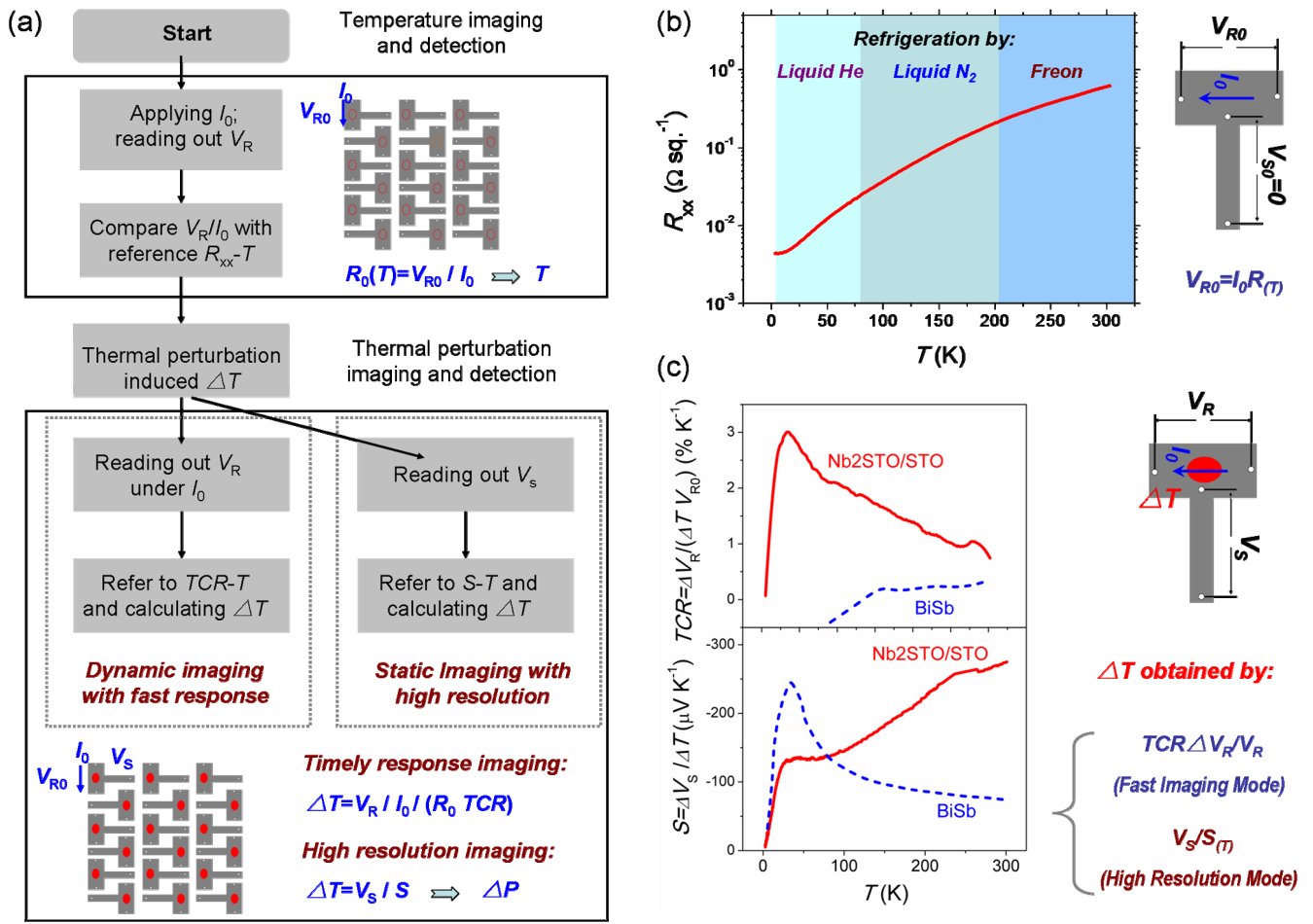

**Figure 4. (a)** The flow chart of the detection strategy as Joule sensor for detections of temperatures and thermal perturbations when combing the thermistor and thermoelectric functionalities. Representative temperature dependence in **(b)** sheet resistance, **(c)** temperature coefficient of resistance and Seebeck coefficient achieved in the presently grown $SrNb_xTi_{1-x}O_3/SrTiO_3$ compared to conventional low temperature thermoelectric material of BiSb [36].